# Exact scattering theory for any straight reflectors in two dimensions.

J.H.Hannay and A.Thain
H.H.Wills Physics Laboratory
University of Bristol
Tyndall Avenue
Bristol BS8 1TL
UK

*Abstract*

The exact Green function for the scalar wave equation in a plane with any set of perfectly reflecting straight mirrors, which may be joined to form corners, is given as a diffraction scattering series. Instances would be slit diffraction in optics, or the Schrodinger equation inside (or outside) a general polygonal enclosure ('quantum polygon billiards'). The method is based on the seminal 1896 Riemann helicoid surface solution by Sommerfeld for optical diffraction by a single corner. It is generalised to account for multiple scatter by adapting the analysis of Stovicek for a closely related problem: a collection of magnetic flux lines (points) in a plane, the multi-flux Aharonov-Bohm effect. The short wavelength limit is shown to yield the 'geometrical theory of diffraction'. For slit diffraction the exact series is shown to coincide with that of Schwarzschild in 1902.

*Introduction*

In two dimensions, as in three, it is not usually possible to write the Green function for the scalar wave equation $\nabla^2 \psi + k^2 \psi = 0$ in the presence of reflecting obstacles (with boundary conditions zero, $\psi=0$, 'Dirichlet' case, or zero normal derivative 'Neumann' case) except as the solution of a boundary integral equation. The iterative solution of this could be considered a scattering series solution and was pioneered by Balian and Bloch [1] (with application to the spectra of quantum 'billiard' enclosures). However if the reflectors are straight line mirrors, then following Thomas Young two hundred years ago, it is tempting to consider the diffraction as arising just from the ends or corners (an end is a $2\pi$ corner) of the mirrors. Although various approximations to this scheme have been successfully pursued, it seems worthwhile to present here the exact scattering series theory underlying the scheme despite its limited practicality.

Wave reflection from straight mirrors invites the introduction of an image waves. A point source of waves in the presence of an infinite straight mirror without ends induces an image source and together they supply the exact wave field. Similarly a point source in the presence of an ideal corner (two half infinite mirrors forming a V shape), admits an exact wave solution obtained in the seminal 1896 paper by Sommerfeld [2][3]. The solution can again be interpreted in terms of images of the source (themselves imaged, generally, carrying on round the corner point forever). With more corners, or mirrors with ends, the challenge is to combine these two parts, mirror images and corner images, or corner diffraction as it would more usually be called. The approximate combination of these has a substantial history.

As a short wave limit one may, in the crudest approximation, ignore corner diffraction effects altogether and use geometrical optics. The light rays either go directly from source to observation point or are reflected specularly (symmetrically) off mirror faces on the way.  Endowed with phase (viewed, for example, as the stationary phase evaluation of a Feynman path integral) the rays combine to exhibit optical interference.  As an improvement to this approximation one can include corner diffraction, treating it as a separate, additive effect.  Most simply this corner diffraction too can be treated using the stationary phase approximation of the Sommerfeld corner solution integral.  This is the strategy in the 'Geometrical theory of diffraction' of Keller [4]. Alternatively, the superior uniform approximation dating back to Pauli can be used [5].  Indeed this is necessary for finding the wave field near the 'shadow boundary' of the geometrically illuminated region where the stationary phase solution diverges falsely.  To cope with a sequence of such tricky marginal scatters the Fresnel diffraction approach of Bogomolny et al [6] would be appropriate.

The exact scattering series will be quite simple to state (in the section on multiple scatter below).  It derives from the Sommerfeld solution for optical diffraction by a single corner which is reviewed in the next section. This is generalised to account for multiple scatter by adapting the analysis of Stovicek [7],[8] for a closely related problem: a collection of magnetic flux lines (points) in a plane, the multi-flux Aharonov-Bohm effect (appendix D).  Corners with internal angle dividing $\pi$ exactly (e.g. a right angle) are analogous to integer flux lines and give zero scatter.  Only genuinely 'diffracting corners' with angles not dividing $\pi$ exactly need feature in the scattering series.  As a partial check on the scattering series it is compared with the 1902 exact scattering series of Schwarzschild [9] for slit diffraction, and shown, with some effort, to agree term by term (appendix C). Another check is its agreement with the geometrical theory of diffraction in the short wavelength limit (appendix B).  Adaptation of the present scattering series to the spectra of 'billiard' enclosures, where a series directly for the trace of the Green function is required, is in progress [10].

*Single scatter*
The famous Sommerfeld solution of 1896 [2],[3] provides the exact Green function for a half infinite straight mirror in two dimensions (fig 1).  Since there is only one diffracting 'corner' involved (the mirror end point) this is single scatter. From single scatter, in the next section, the multiple scatter for more than one diffracting corner is to be built up.  Sommerfeld solves the half infinite mirror problem by what amounts to a method of images trick applied to a simpler system. The simpler system has no mirror at all, and hence no boundary condition (except the usual outgoing one at infinity).  He finds the Green function, instead, for a 'Riemann helicoid surface' - a flattened helicoid comprising an infinite stack of plane layers each joined to its two neighbours in the manner of a screw or spiral staircase (fig 2). To describe his method it is simplest to start, as he does, with a source point at infinite distance producing, therefore, a plane wave on a single layer incident on the vertex (the centre of the helicoid).  The question then is how the wave diffracts and winds around the vertex onto all the other layers. (Actually Sommerfeld starts with just two layers, applicable for the single half mirror, but needs more later for the wedge corner formed by two.  To make the image method more explicit and general, we prefer here to introduce a full helicoid straightaway.)

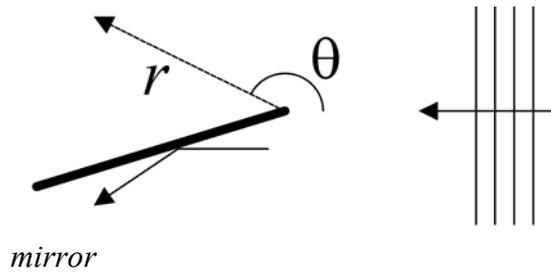

Fig 1. Plane wave hitting a half infinite mirror

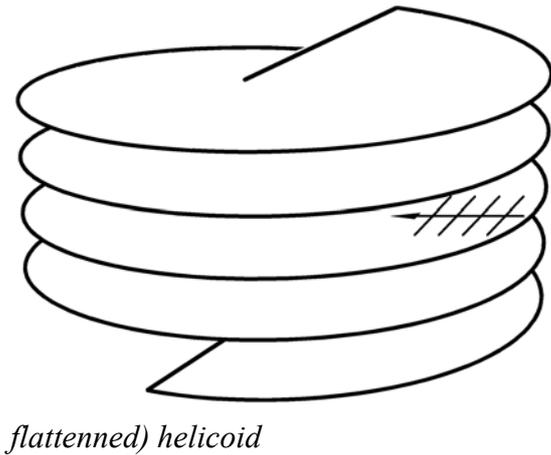

Fig 2. Plane wave on a Riemann (i.e. flattenned) helicoid

To find out he merely writes down the original plane wave as an infinitesimal loop contour integral in the complex plane of the angle coordinate, and then inflates the contour as much as possible (fig 3).

$$\exp(-ikr\cos\theta) = \frac{1}{2\pi i} \oint \frac{1}{\theta' - \theta} \exp(-ikr\cos\theta')\, d\theta' \qquad (1)$$

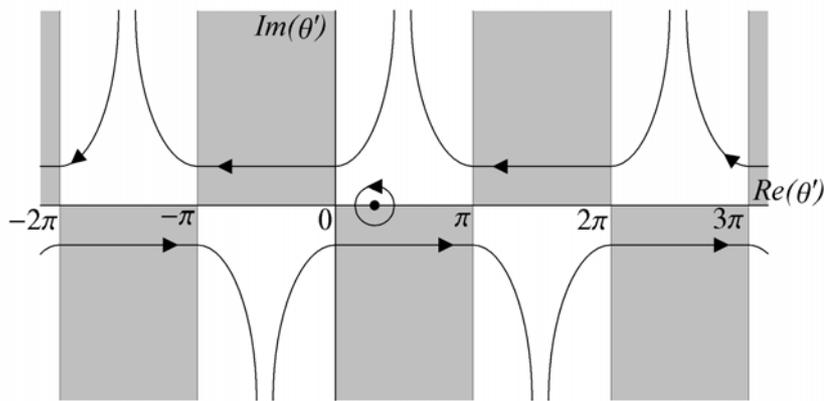

Fig 3. Circle inflates to U shapes

The cleverness is that the fully inflated contour shape has infinitely many similar up and down *U* shaped segments centred at higher and higher angles (period $2\pi$). Consider just a particular pair of U shaped segments, one up, one down, namely those closest to the origin. The integral over this pair of contours alone, with the rest discarded, has a natural interpretation: it describes the wave field on a single layer of the helicoid surface. If $-\pi<\theta<\pi$ it gives the wave field on the geometrically illuminated layer, but if $\theta$ is considered free to range everywhere, $-\infty<\theta<\infty$, the field on other layers is described. To justify this is one checks firstly that the integral obeys the wave equation. This is shown by a shift of the origin of $\theta'$ to $\theta$ so that the integrand as a function of $r, \theta$ then obviously obeys the wave equation. This suffices: the shift makes the contour $\theta$ dependent, and therefore subject to the differentiations in the wave equation, but by analyticity the integral is contour independent, its ends lying in 'valleys' at infinity. Also the integral has the correct singularity on the geometrically illuminated layer (see below) and is outgoing at infinite radius (having subtracted the incident wave on the illuminated layer). Added together the wave fields on all the layers reconstruct the original plane wave as they must.

The wave field on the helicoid is supplied, then, by a U shaped contour integral, that symmetric about $\theta'=-\pi/2$ together with its inverted U partner symmetric about $\theta'=+\pi/2$. These can be shifted sideways somewhat, and rendered rectangular (fig 4) so that the horizontal segments cancel and all that remains is two straight vertical lines. Only if the observation point lies in the geometrically illuminated layer ($-\pi<\theta<\pi$) is there an exception, namely that the horizontal segments do not fully cancel because they go opposite sides of the pole. They yield, instead, the undisturbed plane wave.

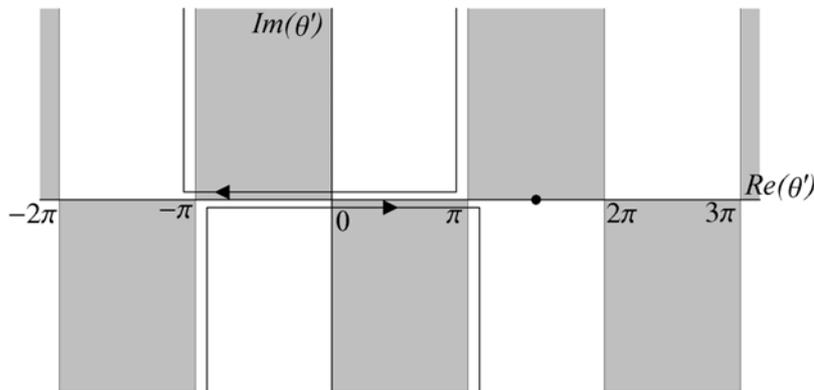

Fig 4. Single U contour and partner (rendered rectangular). The observation point is here shown sited not on the geometrically illuminated layer but on the first shadowed layer of the helicoid.

Thus the wave field on the helicoid ($-\infty<\theta<\infty$) is

$$\text{Step}(\pi-|\theta|)\exp(-ikr\cos\theta)+ \frac{1}{2\pi}\int_{-\infty}^{\infty}ds\ \exp(ikr\cosh s)\left[\frac{1}{\theta+is-\pi}-\frac{1}{\theta+is+\pi}\right] \quad (2)$$

Where the imaginary part of the angle involved in the integration along the vertical lines has been denoted by $s$. Although the first term, corresponding to geometrical illumination, has a line of discontinuity along the 'shadow boundaries' $\theta=\pi$ and $\theta=-\pi$, the second term, the diffraction integral, has a cancelling discontinuity along these lines ensuring that the total wave field is continuous. The cancellation can be checked by noting the residue -iexp(i$kr$) at the pole at $\theta=\pi$ or iexp(i$kr$) at $\theta=-\pi$. This residue (times $2\pi i/2\pi$) supplies the discontinuity in the value of the integral across the lines.

To generate the wave field in the presence of the half infinite mirror it is merely necessary to place image sources on the helicoid. The angles at which they should be placed are found from an angular version of the method of images of an object between two mirrors in one dimension (fig 5). The separation of the two mirrors is $2\pi$ (along the angle axis) with physical space lying between them and containing the true point source. Both the source and the mirrors are imaged out to infinity (in angle).

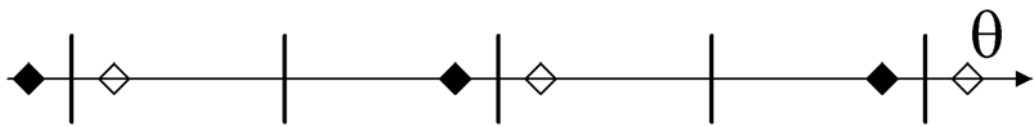

*Fig 5. The angles of plane wave sources. The middle white source is the true one and the others are images. The mirrors on either side of the true source represent the two faces of the true half infinite mirror. For Dirichlet boundary conditions alternate image sources are negative (shown black), while for Neumann ones all are positive.*

The summation over all the white sources of fig 5 (separation $4\pi$) is straightforward using the identity $\Sigma(n+x)^{-1}=\pi\cot(\pi x)$, over $-\infty<n<\infty$. Likewise for the black sources. The summation can be made explicit at any desired stage, but we proceed, for simplicity with the unsummed helicoid, remarking only that for the plane wave source under consideration the result of the summation can be manipulated into the famous exact Fresnel integral form.

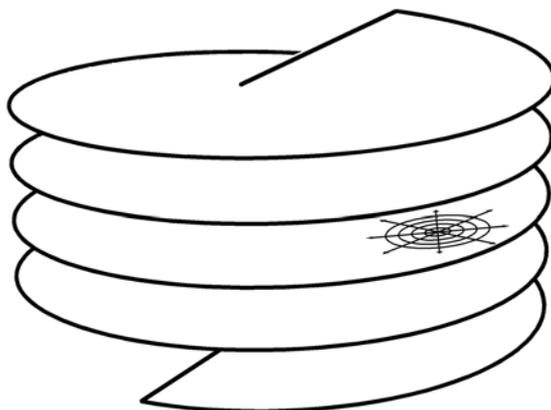

*Fig 6. Point source on a Riemann helicoid*

Sommerfeld proceeded to find the wave field from a source at finite distance $r_0$, that is, he found the Green function (fig 6) for the helicoid. The procedure is simply to replace the plane wave in the integral (1) by the free space Green function, namely the Hankel function $\frac{1}{2i}H_0^{(1)}(k\,|\mathbf{r}-\mathbf{r_0}|)$ (which solves $\frac{1}{2}\nabla^2\psi+\frac{1}{2}k^2\psi=-\delta(\mathbf{r}-\mathbf{r_0})$ in two dimensions, the constants being appropriate to quantum mechanics with $\hbar=m=1$). Measuring the angle $\theta$ from the direction of the source as before, the Green function for the helicoid is

$$\text{Step}(\pi-|\theta|) \tfrac{1}{2i} H_0^{(1)}(k\sqrt{r_0^2 + r^2 - 2rr_0 \cos\theta}) +$$

$$\int_{-\infty}^{\infty} ds \; \tfrac{1}{4\pi i} H_0^{(1)}(k\sqrt{r_0^2 + r^2 + 2rr_0 \cosh s}) \left[\frac{1}{\theta + is - \pi} - \frac{1}{\theta + is + \pi}\right] \quad (3)$$

One notes that the Hankel function in the 'geometrical illumination' term contains the real angle $\theta$, while that inside the diffraction integral contains the imaginary part of the angle, $s$. The angular diffraction or scattering factor in square brackets contains the full complex angle $\theta+is$

$$\left[\frac{1}{\theta + is - \pi} - \frac{1}{\theta + is + \pi}\right] \quad (4)$$

Once again the value of the diffraction integral has a line of discontinuity on $\theta=\pi$ and $\theta=-\pi$ with the jump in value obtained from the residue at the corresponding pole. This cancels the discontinuity in the geometrical term so that the total wave field on the helicoid is continuous. Since the geometrical wave field is zero on the shadow side, one can obtain the value of the wave field exactly on the shadow boundary by taking the limit of the diffracted wave field from that side. A similar principle applies to the case of more than one scatter.

For the purpose of concatenation of scatters (linking them successively into a chain) in the next section it is worth re-expressing this formula for the Green function in terms of 'time'. For example the geometrical illumination term becomes

$$\tfrac{1}{2i} H_0^{(1)}(k \times distance) = \tfrac{-1}{2\pi} \int_0^\infty \exp(\tfrac{1}{2}ik^2 t) \; \exp(\frac{i}{2t} distance^2) \; \frac{dt}{t} \quad (5)$$

This integral representation of the Hankel function has the physical interpretation of the Fourier transform of the standard time propagator in free space. With $\hbar = m = 1$ this propagator is

$$K_0(distance, t) \equiv \frac{1}{2\pi i t} \exp\left(\frac{i \, distance^2}{2t}\right) \quad (6)$$

and has the form $(1/time) \exp(i \times kinetic\ energy \times time)$. In full, the new expression for the Green function on the helicoid is

$$\text{Step}(\pi-|\theta|) \tfrac{-1}{2\pi} \int_0^\infty \exp(\tfrac{1}{2}ik^2 t) \; \exp(\frac{i}{2t}(r_0^2 + r^2 - 2rr_0 \cos\theta)) \; \frac{dt}{t} +$$

$$\tfrac{-1}{4\pi^2} \int_0^\infty \frac{dt}{t} \int_{-\infty}^\infty ds \; \exp(\tfrac{1}{2}ik^2 t) \; \exp(\frac{i}{2t}(r_0^2 + r^2 + 2rr_0 \cosh s)) \left[\frac{1}{\theta + is - \pi} - \frac{1}{\theta + is + \pi}\right] \quad (7)$$

The final step, following Stovicek [7], is to make more explicit the separation into before scatter (subscript 0) and after scatter (subscript 1). First, trivially, $r$ will be renamed $r_1$. Then new integration variables $t_0$ and $t_1$ will be introduced by making the substitution $s=\ln(r_1t_0/r_0t_1)$, with the relation $t_0+t_1=t$. The Jacobian of this transformation $s,t \to t_0,t_1$ is $(t/t_0t_1)$. Natural interpretations for $t_0$ and $t_1$ are the 'time' durations before and after scatter and this is suggestive for the later step of concatenating scatters so we adopt this 'time' nomenclature henceforth. Actually it is worth retaining the $t$ integral unevaluated, easy though it is, for later comparison. The Green function for the helicoid thus becomes

$$\text{Step}(\pi-|\theta|)\frac{-1}{2\pi}\int_0^\infty \exp(\tfrac{1}{2}ik^2 t)\, \exp(\frac{i}{2t}(r_0^2+r_1^2-2r_1r_0\cos\theta))\,\frac{dt}{t} +$$

$$\frac{-1}{4\pi^2}\int_0^\infty\int_0^\infty\int_0^\infty \frac{dt_0}{t_0}\frac{dt_1}{t_1}dt\,\delta(t_0+t_1-t)\,\exp(\tfrac{1}{2}ik^2t)\,\exp(\frac{ir_0^2}{2t_0})\,\exp(\frac{ir_1^2}{2t_1}) \times$$

$$\left[\frac{1}{\theta+i\ln(r_1t_0/r_0t_1)-\pi} - \frac{1}{\theta+i\ln(r_1t_0/r_0t_1)+\pi}\right] \qquad (8)$$

This new expression has the merit of having decomposed the diffraction integrand into a form: free propagation, then scatter, then free propagation again. (Performing the $t$ integral would make that completely explicit). If the scatter term looks ugly, one may note that the complex angle $\theta+is$ has nevertheless a certain elegance. Temporarily thinking of the two dimensional space of propagation as a complex plane with the source as drawn on the real axis, $\theta+is$ is the change of the logarithm of the complex velocity at the scatter. If one denotes initial velocity by the real number $r_0/t_0$, and final velocity by $(r_1/t_1)\exp i\theta$, then
$\theta+is=\ln((r_1/t_1)\exp i\theta) - \ln(r_0/t_0)$.

By virtue of the decomposition just mentioned the form (8) allows concatenation as we learn from Stovicek [7], [8]; we can generalise to a chain of scatters in the next section. Before doing so, it is useful, finally to revert to Sommerfeld and present the summed formula, the Green function for the half infinite mirror or indeed a reflecting wedge of any positive open angle $\gamma$ (the half screen being a wedge of open angle of $2\pi$). This, likewise, will be concatenable; as indicated before, summation and concatenation are independent procedures. The results will then describe the Green function for any arrangement of straight mirrors perhaps joined to form corners or polygons.

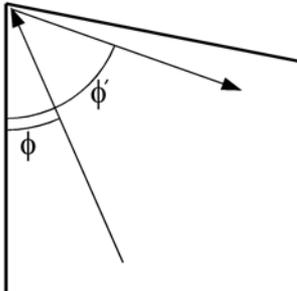

*Fig 7. Defining the angles for corner scatter*

The arrangement of image source angles for a corner of open angle $\gamma$ is just as in fig 5 above except that the gap between the mirrors is now $\gamma$ rather than $2\pi$. It is useful to adopt a system of measuring angles which will allow generalisation. At every corner between two mirrors we shall measure angles anti-clockwise from the clockwise extreme of the corner opening (fig 7). The two joining straight mirrors thus have angles zero and $\gamma$. The incoming direction, or rather its reverse, will be denoted by the angle $\phi$, and the outgoing direction to the field point by $\phi'$. Thus both angles lie between zero and $\gamma$. Pictorially one can represent an individual image term in this summation in a purely schematic way as a 'rattling' or swinging and bouncing (like the clapper of a bell) between the two mirrors (fig 8). We shall use the word 'rattle' to describe this angular reflection process. In keeping with the description 'diffraction', rattle reflections lead to an outgoing direction from a corner which is not determined by the incoming one, in contrast to ordinary specular reflections on a mirror face.

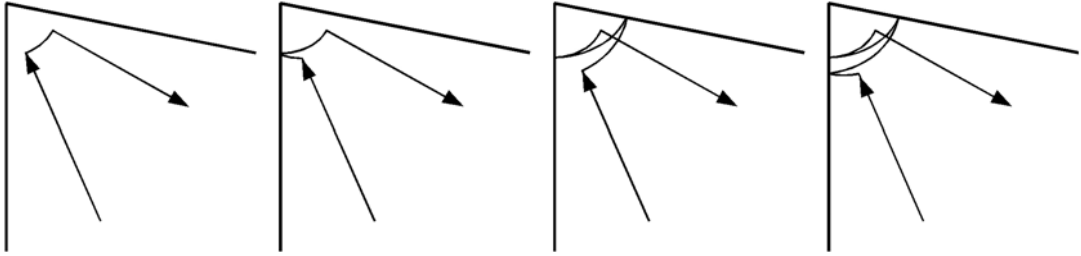

*Fig 8. The rattles with reflection number M=0, -1, +2, -3*

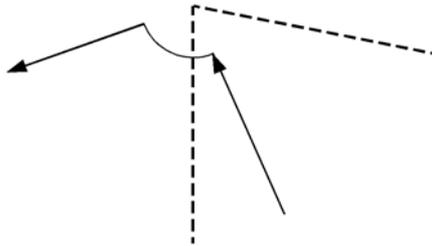

*Fig 9. Unfolded version of M=-1 rattle*

The various scattering angles from the rattlings are the total angles of swing (signed according to whether their initial swing direction is anticlockwise or clockwise). Thus with the number of reflections $|M|$ (and $-\infty \leq M \leq \infty$), the 'unfolded' scattering angles are $\theta=\phi'-\phi+M\gamma$, for $M$ even or $\theta=-\phi'-\phi+(M+1)\gamma$, for $M$ odd (fig 9). With Dirichlet boundary conditions the even $M$ sum corresponds to the sum over positive images and the odd $M$ sum to the sum over negative images (for Neumann all are positive).

The Sommerfeld Green function for the wedge is then the sum of direct free space Green functions from (possibly multiple) specular reflections on mirrors plus the single scatter diffracted ray Green function, namely

$$\frac{-1}{4\pi^2} \int_0^\infty dt \int_{-\infty}^\infty ds \; \exp(\tfrac{1}{2}ik^2 t) \; \exp(\frac{i}{2t}(r_0^2 + r^2 + 2rr_0 \cosh s)) \; \{f(\phi'-\phi) - f(-\phi'-\phi)\} \quad (9)$$

where the $f$ is the sum of primitive scattering factors (4) over rattles using $\Sigma(n+x)^{-1}=\pi\cot(\pi x)$,

$$f(\bullet) \equiv \frac{\pi}{2\gamma}\left[\cot\frac{\pi(\bullet + is - \pi)}{2\gamma} - \cot\frac{\pi(\bullet + is + \pi)}{2\gamma}\right] \qquad (10)$$

The first *f* in (9) describes the sum over even *M* images, the second that over odd *M* ones, and the negative sign between them is appropriate for Dirichlet boundary conditions (for Neumann it would be positive). The sign between the *cot* functions in (10) is always negative, as in (4). One may note, as mentioned in the introduction, that if $\gamma$ divides $\pi$ then *f*=0 by the $\pi$ periodicity of the function *cot* and there is no corner diffraction.

The diffracted wave field (9) is continuous except on two 'shadow boundaries' across which it has a discontinuity. This exactly cancels the accompanying discontinuity in the geometrical illumination field, so that the total wave field has no discontinuity. The discontinuity arises mathematically because, in moving the field point across the shadow boundary, a pole in *f* crosses the (real *s*) axis of integration at the origin. This association (of geometrical shadow and diffraction pole at the origin) comes about as follows.

Consider an ordinary ray (not corner diffracted) from the source which runs along side (indefinitely close to) the ray coming directly in to the corner. Each of the two possibilities for which side it runs is to give rise to a shadow boundary. The ordinary ray (choosing one possibility) will just miss the corner and instead reflect specularly off the two mirrors alternately some number of times (0,1,2…) close to the corner before emerging from the corner region (radially in the limit). It constitutes, by definition, a shadow boundary. We can now argue that if the outgoing diffracted ray (at angle $\phi'$) coincides with this shadow boundary there is a pole in an *f*. Among all the different rattle paths there will be one whose reflection sequence matches that of the ordinary ray. When unfolded, the ordinary ray is rendered straight. If the matching rattle path were likewise unfolded, it too would emerge in a straight ahead direction. It has an unfolded scattering angle $\theta$ (as defined above) equal to $\pm\pi$, and therefore its angular scatter factor (4) has a pole at the origin in *s*, which is inherited by the sum (10) over rattles, *f*.

*Multiple scatter*

The Green function is a sum of contributions from all possible geometrical (undiffracted) rays, and from all possible diffracted rays joining the source point to the observation point via a sequence of corners. A diffracted ray with *n* corner scatters (i.e. diffractions) has *n*+1 'legs'. Any leg may involve one or more specular reflections off mirror faces (it is still all one leg). Successive corners must be visible from each other (possibly via specular reflections). Successive corners which lie on the same mirror (at opposite ends of it) are considered as mutually visible, with the ray travelling alongside the mirror surface – a 'surface leg' (which, however will be shown to count half). 'Accidental' contact of a ray with a corner is not allowed. That is, if during its journey, a ray leg happens to reflect specularly off a mirror at zero distance from a corner, this corner must be counted as a corner in the scattering sequence; the ray leg must be split into two consecutive legs. The scatter at this corner is necessarily of the shadow boundary type just discussed for single scatter. Although non-generic (because it arose only in special accidental circumstances) it is particularly important when the Green function is traced [6][10]. We return to this shadow boundary case in appendix A.

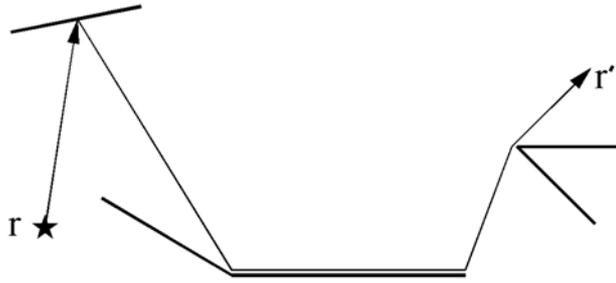

*Fig 10. A diffractive ray with three corner scatters, and therefore four legs in all. The first leg involves a specular reflection and the second is a 'surface' leg.*

The form of the Green function is a concatenation of scatters (fig 10): leg, scatter, leg, scatter, …leg, scatter, leg. However these stages are linked to each other (in contrast with approximate treatments): they are part of a multiple integral, one integration for each scatter. The justification of the formula lies in the discovery of Stovicek [7] [8] that the exact Green function on the multi-helicoid 'covering surface' is a concatenation of scatters (appendix D). He used this for the multiple flux line Aharonov-Bohm effect. Our use of it, for our system of mirrors rather than flux lines, is simpler in that there are no magnetic flux phase factors, but more complicated in that some book keeping is necessary to account correctly for images and parity changes. Surface rays, particularly, require careful accounting.

The basic principle for forming the Green function is easy to state (though infinitely uneconomical in the number of terms prior to simplification): every distinct diffracted ray (with every different number of internal rattles at corners) needs counting once. Its contribution is found by first unfolding the ray path fully and assigning positive durations to the legs. Then form the chain $iK_0\times[\ ]\times iK_0\times[\ ]\times\ldots\times[\ ]\times iK_0$ where the $K_0$ factors (6) have their different leg lengths and durations, and $[\ ]$ stands for the different primitive corner diffraction factors (4) with $s=\ln$(speed after/speed before). Finally integrate this product over all leg durations (with the Fourier transform factor $(-i)\ \exp[i\frac{1}{2}k^2\times$ total duration$]$). For Dirichlet boundary conditions an additional factor $-1$ is needed for every reflection (rattle or specular). Both the scatter term product and the leg term product and can be simplified separately, first the former and then the latter (starting with the paragraph of eqn 13).

One can fix attention on a particular diffracted ray path (sequence of legs between scatter corners) and evaluate the rattle summation at each corner. As before (9), for a given corner, there are separate results $f$ for even and odd numbers of rattle reflections there. However with more than one scatter there is a subtlety and it is not correct merely to take the difference (or sum) of even and odd parts for each corner and form their product. Instead, as explained below, the total scattering factor will be expressed in terms of a product of (2×2) matrices.

The subtlety mentioned concerns parity. The unfolding process for a ray reverses the parity of space at each unfold (i.e. reflection, whether it is a rattle reflection or a specular one). Thus the positive angles $\phi'$ and $\phi$, can, on the unfolded ray, acquire negative sense. This means that for more than one scatter there are four, rather than two, possible arguments for $f$. For even rattle number one has $f(\pm(\phi'-\phi))$ and for odd $f(\pm(-\phi'-\phi))$, where the parity $\pm$ is determined by whether the total number of reflections of any kind *prior to* the corner in question is even or odd. For the $j^{th}$ scatter corner denote by $f_j(\bullet)$ the form (10) with its non-explicit arguments, namely the quantities $s$ and $\gamma$, each having subscript $j$. The four possibilities are usefully assembled into a (2×2) corner scatter matrix for the $j^{th}$ scatter corner,

$$F_j \equiv \begin{pmatrix} f_j(\phi'_j - \phi_j) & f_j(\phi'_j + \phi_j) \\ f_j(-\phi'_j - \phi_j) & f_j(-\phi'_j + \phi_j) \end{pmatrix} \qquad (11)$$

It will be shown that the correct total scattering factor is given in terms of the string of these $n$ matrices (in reverse order) interlaced with leg ones, $\begin{pmatrix} 1 & 0 \\ 0 & 1 \end{pmatrix}$ for a leg with no specular reflections (or an even number of them), or $\begin{pmatrix} 0 & 1 \\ 1 & 0 \end{pmatrix}$ for an odd number of them. For a surface ray leg the matrix is $\begin{pmatrix} 1/2 & 0 \\ 0 & 1/2 \end{pmatrix}$. For example the resulting matrix for fig 10 is $\begin{pmatrix} 1 & 0 \\ 0 & 1 \end{pmatrix} \times F_3 \times \begin{pmatrix} 1 & 0 \\ 0 & 1 \end{pmatrix} \times F_2 \times \begin{pmatrix} 1/2 & 0 \\ 0 & 1/2 \end{pmatrix} \times F_1 \times \begin{pmatrix} 0 & 1 \\ 1 & 0 \end{pmatrix}$. With this understanding for the leg matrices, the total scattering factor is given for Dirichlet boundary conditions by

$$(1 \ -1) \ [(\ ) \times F_n \times (\ ) \times \ldots \times (\ ) \times F_2 \times (\ ) \times F_1 \times (\ )] \begin{pmatrix} 1 \\ 0 \end{pmatrix} \qquad (12)$$

For Neumann boundary conditions the vector (1, -1) is replaced by (1, 1).

The justification of this matrix representation is as follows. The vector $\begin{pmatrix} 1 \\ 0 \end{pmatrix}$ input on the right hand side of the product signifies the initial, real space, parity. If there is no specular reflection before the first corner scatter (or if there is an even number of them) the initial leg matrix is $\begin{pmatrix} 1 & 0 \\ 0 & 1 \end{pmatrix}$ and parity is retained on entering the scatter, otherwise it is reversed by the initial leg matrix $\begin{pmatrix} 0 & 1 \\ 1 & 0 \end{pmatrix}$ to $\begin{pmatrix} 0 \\ 1 \end{pmatrix}$. Muliplication by $F_1$ yields a vector with components corresponding to the true and reversed parity conditions exiting the corner, namely (for even prior specular reflections) ($f_1(\phi_1'-\phi_1)$, $f_1(-\phi_1'-\phi_1)$). For odd prior specular reflections instead, ($f_1(-\phi_1'+\phi_1), f_1(\phi_1'+\phi_1)$) would be produced. The parity condition information is thus retained through the matrix product and the true and reversed parity results are finally combined with a sign difference for Dirichlet boundary conditions or without for Neumann.

Surface ray legs require further discussion to justify their associated matrix $\begin{pmatrix} 1/2 & 0 \\ 0 & 1/2 \end{pmatrix}$. This is half times the matrix for zero (or even) specular reflections, and both the 'half' and the 'zero' need explanation. Taking zero for granted, temporarily, the need for the factor half becomes clear as follows by examining the rattles at either end of the mirror (fig 11).

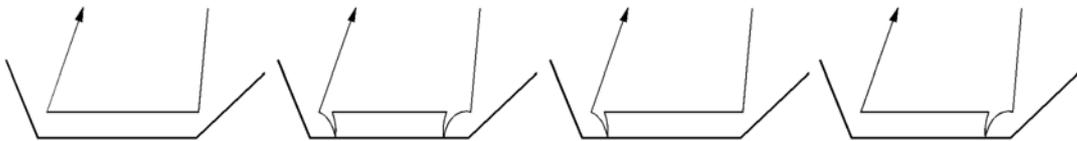

*Fig 11. Possibilities for surface leg rattles. Only two, not four should be counted. For simplicity the case shown has no additional rattle reflections on the side mirrors.*

There can be a rattle reflection at neither end or at both ends, or at one and not the other (two choices), and these are related in a rather special way. Unfolding would normally distinguish paths with different rattlings, but of these four possibilities, even when unfolded, the first two are essentially identical paths. They have identical contributions to the scattering sum and ought not be counted as distinct. The same applies to the latter pair. Such duplicates should be counted just once, not twice, and the easiest prescription is to count them all and divide by two. Returning to the question of 'zero' specular reflections assumed above, actually this assumption is not required. The matrix $\begin{pmatrix} 1/2 & 0 \\ 0 & 1/2 \end{pmatrix}$ can be replaced by $\begin{pmatrix} 0 & 1/2 \\ 1/2 & 0 \end{pmatrix}$ associated with a specular reflection on the surface. There is no change in the result because, as a surface ray, the leg either has zero exit angle from the first corner, or zero entry angle to the second. (In fig 11 the latter holds by the anticlockwise definition of angles, but flipping the pictures upside down would give the former). This means, in the former case, that the first corner matrix $F$ has its two rows identical and therefore $\begin{pmatrix} 1/2 & 0 \\ 0 & 1/2 \end{pmatrix} F = \begin{pmatrix} 0 & 1/2 \\ 1/2 & 0 \end{pmatrix} F$, or in the latter case the second corner $F$ has its two columns identical so that $F \begin{pmatrix} 1/2 & 0 \\ 0 & 1/2 \end{pmatrix} = F \begin{pmatrix} 0 & 1/2 \\ 1/2 & 0 \end{pmatrix}$, demonstrating the equivalence. Using either matrix ensures that all truly distinct paths are counted once.

The Green function for the system is then given by a sum over any number of scatters $n$ from one to infinity. An $n=0$ term is included for neatness, referring to geometrical illumination (possibly via specular reflections) with no integrations and no scatters, just a single free propagator $K_0$ (6) for each geometrical ray. Denoting the source point by **r**, the observation point by **r'**, and the energy by $E \equiv \frac{1}{2}k^2$ (with $\hbar = m = 1$), the Green function is

$$G(\mathbf{r},\mathbf{r'},E) = \frac{1}{i} \int_0^\infty K(\mathbf{r},\mathbf{r'},t) \exp(\tfrac{1}{2}ik^2 t)\, dt, \tag{13}$$

$$K(\mathbf{r},\mathbf{r'},t) = \sum_{n=0}^\infty i^n \sum_{\substack{n \text{ vertex} \\ \text{paths}}} \int_0^\infty dt_0 dt_1 ... dt_n K_0(r_0,t_0) K_0(r_1,t_1)...K_0(r_n,t_n)\, \delta(t_0 + t_1 ... + t_n - t) \times$$

$$(1 \;\; -1)\; [(\;)\times F_n \times (\;) \times ... \times (\;) \times F_2 \times (\;) \times F_1 \times (\;)] \begin{pmatrix} 1 \\ 0 \end{pmatrix} \tag{14}$$

for Dirichlet boundary conditions. For Neumann boundary conditions the vector (1, -1) is replaced by (1, 1).

The $n+1$ integrations reduce by virtue of the $\delta$-function to $n$ integrations over the ratio variables associated with each corner, or of their logarithms, the variables $s$, as in eqn(8). Stovicek provides the required Jacobian relation

$$ds_1 ds_2 ... ds_n = t \delta(t_0 + t_1 + ... + t_n - t) \frac{dt_0}{t_0} \frac{dt_1}{t_1} ... \frac{dt_n}{t_n} \tag{15}$$

The sum of the exponents of the $K_0$ factors is straightforwardly converted:

$$\frac{i}{2}\left( \frac{r_0^2}{t_0} + \frac{r_1^2}{t_1} + ... + \frac{r_n^2}{t_n} \right) \tag{16}$$

$$= \frac{i}{2t}(r_0 + r_1 e^{s_1} + r_2 e^{s_1+s_2} .. + r_n e^{s_1+s_2...+s_n})(r_0 + r_1 e^{-s_1} + r_2 e^{-s_1-s_2} .. + r_n e^{-s_1-s_2...-s_n}) \quad (17)$$

$$= \frac{i}{2t}\left[\sum_{0 \le j \le n} r_j^2 + 2 \sum_{0 \le j < m \le n} r_j r_m \cosh(s_{j+1}... + s_m)\right] \quad (18)$$

$$\equiv \frac{i}{2t} R^2(s_1, s_2..., s_n) \quad (19)$$

defining the quantity $R$ to shorten subsequent formulas.

The propagator is therefore:

$K(\mathbf{r},\mathbf{r}',t)=$

$$\sum_{n=0}^{\infty} (\tfrac{1}{2\pi})^n \sum_{\substack{n \text{ vertex} \\ \text{paths}}} \int_{-\infty}^{\infty} ds_1 ds_2 ... ds_n \frac{1}{2\pi i t} \exp\left[\frac{i}{2t} R^2(s_1, s_2, ...s_n)\right] \times$$

$$(1 \;\; -1) [(\;) \times F_n \times (\;) \times ... \times (\;) \times F_2 \times (\;) \times F_1 \times (\;)] \binom{1}{0} \quad (20)$$

for Dirichlet boundary conditions. For Neumann boundary conditions the vector (1, -1) is replaced by (1, 1).

Finally this is easy to Fourier transform with respect to $t$ (setting $K=0$ for $t<0$) to obtain the main result, the Green function:

$G(\mathbf{r},\mathbf{r}',E)=$

$$= \sum_{n=0}^{\infty} (\tfrac{1}{2\pi})^n \sum_{\substack{n \text{ vertex} \\ \text{paths}}} \tfrac{1}{2i} \int_{-\infty}^{\infty} ds_1 ds_2 ... ds_n H_0^{(1)}[kR(s_1, s_2, ...s_n)] \times$$

$$(1 \;\; -1) [(\;) \times F_n \times (\;) \times ... \times (\;) \times F_2 \times (\;) \times F_1 \times (\;)] \binom{1}{0} \quad (21)$$

for Dirichlet boundary conditions. For Neumann boundary conditions the vector (1, -1) is replaced by (1, 1). Once again, the zero scatter $n=0$ term is to be understood as having no integrations and no scattering factor matrices $F$. It is a summation over all 'direct' ray paths from $\mathbf{r}$ to $\mathbf{r}'$ (with possible specular reflections) in which $R$ is taken equal to $r_0$, the total length of the ray path in question.

**Appendix A. Shadow boundaries**

For the observation point lying on a shadow boundary of a ray that has had multiple scatters there are cancelling discontinuities. The cancellation (appendix D) is similar to that described for single scatter. One discontinuity arises from a pole in the diffraction integral at the scatter casting the shadow (the final one), and the other from the scatter immediately before, with the final one omitted (analogous to the geometrical, source wave field in single scatter).

The 'accidental' circumstance of a shadow boundary arising in the course of a scatter path rather than at the end, i.e. one scattering corner lying exactly on the shadow boundary of another, is dealt with similarly. The circumstance shows up mathematically as a pole at the origin in the $s$ integral associated with the scatter casting the shadow. The prescription for which side of the pole the integration contour should pass emerges as follows from an investigation of the geometry responsible for the singularity. One particular rattle ray at the corner casting the shadow unfolds to emerge straight ahead after bypassing the corner, making one of its two terms (4) singular at *s*=0. The ±π sense of the bypass determines the prescription for the integration contour: the contour should bypass the pole in the complex *s* plane in the sense opposite to that with which the physical unfolded rattle ray bypasses the corner. To justify this one first notes that hypothetical infinitesimal transverse shifts in physical and complex planes match in sign: a rightward transverse shift in the corner position, looking along the ray, causes a positive increment in scattering angle which shifts the pole in (4) downwards (i.e. rightwards looking along the path) in the complex plane. Now consider a shift of the corner position in a direction *away* from its bypass semicircle. This corresponds to allowing geometrical visibility passed the corner. For the integration contour, allowing geometrical visibility means acquiring a pole residue by pushing the pole *towards* the bypass loop of the contour. The two bypasses must therefore be in opposite senses.

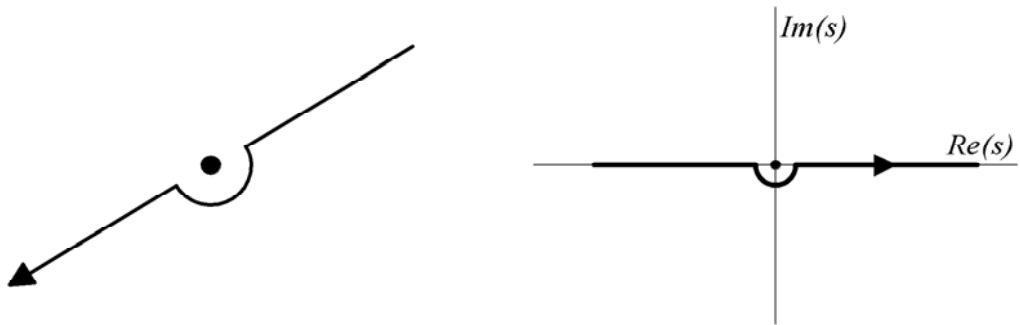

*Fig 12. Left: An unfolded shadow boundary rattle ray bypassing its corner point with an angle –π. Right: the corresponding integration contour for the s integral for that corner bypasses its pole with the opposite sense; +π.*

In the even more special case that there are several such accidental shadow boundary scatters in succession the prescription is the natural generalization. There is a single singular rattle ray which unfolds to be straight ahead after bypassing each relevant corner (the ones casting shadows on the next). The sense ±π of each bypass needs to be reproduced, oppositely, in the complex plane contour of its own *s* integral.

**Appendix B. Geometrical theory of diffraction**

The stationary phase evaluation of the Green function will now be shown (in outline) to lead to the so called 'Geometrical theory of diffraction' [4], [12]. This is the short wavelength or semiclassical limit of diffraction theory with different versions appropriate to many different circumstances (e.g. curved mirrors, refracting interfaces etc.). For our circumstance of straight mirrors in two dimensions the most obvious limitations of the geometrical theory (and correspondingly the stationary phase approximation of our Green function) show up close to any shadow boundary, being falsely singular there. Also, for diffracted ray paths with surface ray legs the geometrical theory gives (correctly) zero contribution (for Dirichlet boundary conditions) at the usual order in $k$, as does stationary phase. The evaluation at the next order in $k$ is discussed at the end below.

To obtain the short wavelength limit the Hankel function in (21) is replaced by its asymptotic form (which is equivalent to evaluating the time integral in (5) by stationary phase) $H_0^{(1)}(kR) \approx \sqrt{2/i\pi kR} \exp(ikR)$. The exponent $ikR$ is stationary when $s_1=s_2=\ldots\ldots=s_n=0$ and expanding the *cosh* functions with $L \equiv r_0+r_1\ldots+r_n$,

$$R(s_1, s_2\ldots, s_n) \approx \sqrt{L^2 + \sum_{0 \le j < m \le n} r_j r_m (s_{j+1}\ldots + s_m)^2} \approx L + \frac{1}{2L} \sum_{0 \le j < m \le n} r_j r_m (s_{j+1}\ldots + s_m)^2 \quad (22)$$

There is simplification of the corner scatter product too. Since the functions $f$ in the scattering matrix product are non-singular (excluding the shadow boundaries mentioned) all the $s_j$ in them can be set to zero for the stationary phase evaluation. (This is not true, however, for surface rays in the Dirichlet case which is dealt with later). The functions $f$ (10) are now symmetric functions of their argument ($f(\bullet)=f(-\bullet)$) and therefore each matrix $F$ (11) has its two diagonal terms equal and its two off-diagonal terms equal. That is, $F_j$ decomposes into $a_j \begin{pmatrix} 1 & 0 \\ 0 & 1 \end{pmatrix} + b_j \begin{pmatrix} 0 & 1 \\ 1 & 0 \end{pmatrix}$ with $a_j \equiv f_j(\phi_j'-\phi_j)$ and $b_j \equiv f_j(-\phi_j'-\phi_j)$. Using the fact that the matrices $\begin{pmatrix} 1 & 0 \\ 0 & 1 \end{pmatrix}$ and commute $\begin{pmatrix} 0 & 1 \\ 1 & 0 \end{pmatrix}$ (so the interleaving leg matrices, which have the same form, can be moved away), the product of such $F_j$ matrices can be simplified using the identity

$$\prod \left[ a_j \begin{pmatrix} 1 & 0 \\ 0 & 1 \end{pmatrix} + b_j \begin{pmatrix} 0 & 1 \\ 1 & 0 \end{pmatrix} \right] = \begin{pmatrix} 1 & 0 \\ 0 & 1 \end{pmatrix} \tfrac{1}{2} \left[ \prod (a_j + b_j) + \prod (a_j - b_j) \right] +$$
$$\begin{pmatrix} 0 & 1 \\ 1 & 0 \end{pmatrix} \tfrac{1}{2} \left[ \prod (a_j + b_j) - \prod (a_j - b_j) \right] \quad (23)$$

With the initial and final vectors, (1,0) and (1,-1) for Dirichlet boundary conditions inserted, the whole the scattering factor becomes the product of pairs $[f_j(\phi_j'-\phi_j)-f_j(-\phi_j'-\phi_j)]$ as in (9), with an additional factor $-1$ for each specular reflection. (For Neumann boundary conditions the insertion of (1,1) as the final vector means the two functions $f$ in each pair are added instead of subtracted and the factor $-1$ is absent).

Returning to (22), having argued that the corner scattering product acts a constant as far as the integrations are concerned, one is left with a multiple Gaussian integral. Changing the variables of integration to $s'_1=s_1$, $s'_2=s_1+s_2$, …., $s'_n=s_1+s_2\ldots+s_n$, and defining $s'_0=0$ for notational convenience, we have

$$R(s_1, s_2\ldots, s_n) \approx L + \frac{1}{4L} \sum_{0 \le j \le n} \sum_{0 \le m \le n} r_j r_m (s'_j - s'_m)^2 \quad (24)$$

Then the Gaussian integrations can be performed directly to yield a determinant whose rows and columns can be simplified by inspection

$$\int \exp[ik(R-L)] ds_1 \ldots ds_n \approx \sqrt{\left(\frac{4\pi i L}{k}\right)^n \frac{1}{L^{n-1} r_0 r_1 \ldots r_n}} \quad (25)$$

(More systematically to obtain (25) one can eliminate the off-diagonal determinant elements of (24) by capturing them in a perfect square and introduce an auxiliary integration to represent it: $\int \exp[iku(\Sigma r_m s'_m)/2L + iku^2/8L]\,du = \sqrt{2\pi i}\exp[-ik(\Sigma r_m s'_m)^2/2L]$, and then reverse the order of integration). Thus, in summary, stationary phase evaluation of the Green function yields the zero $s$ value of the integrand of (21) times the integral (25). This result reproduces the 'Geometrical theory of diffraction'.

The special case of surface ray legs can also be investigated in the geometrical limit. For Neumann boundary conditions surface rays are no different from ordinary ray legs (apart form the surface factor one half). For Dirichlet ones however, a surface ray contributes at a lower order in $k$ than a non-surface one. This arises because the special angles involved (zero or $\gamma$ for the angles at the ends of the leg). Using the symmetry of $f$ at $s=0$ ($f(\bullet)=f(-\bullet)$) and also the periodicity of $f$ with period $2\gamma$ it follows that the corner scatter matrices $F_j$ and $F_{j+1}$ at each end of a surface ray leg have all four of their elements equal. In particular $[f_j(\phi_j'-\phi_j)-f_j(-\phi_j'-\phi_j)]=0$ (and similarly with subscripts $j+1$) rendering the product of pair terms over all $j$ zero.

To obtain a non-zero result, the values of $s_j$ and $s_{j+1}$ must not be set to zero. For simplicity we shall assume that this is the only surface leg so that all the other $s$ values can be set to zero. Instead the $f_j$ and $f_{j+1}$ must be Taylor expanded to first order, $f \approx \text{constant} + s\dot{f}$ where the over-dot indicates differentiation with respect to $s$, and their matrices formed: $F_j \approx \text{constant} + s_j \dot{F}_j$ and $F_{j+1} \approx \text{constant} + s_{j+1}\dot{F}_{j+1}$. The constant part of each matrix (with all its four elements equal) can be discarded since its product with the other leads to zero contribution as stated, and also terms linear in $s$ vanish on integration. Due to the antisymmetry ($\dot{f}(\bullet)=-\dot{f}(-\bullet)$) of the $s$ derivative at $s=0$, together with the periodicity of $f$, period $2\gamma$, the matrix $\dot{F}_j$ has the form $\dot{f}_j(\phi_j'-\phi_j)\begin{pmatrix}1 & -1\\ 1 & -1\end{pmatrix}$, and $\dot{F}_{j+1}$ has the form $\dot{f}_{j+1}(\phi_{j+1}'-\phi_{j+1})\begin{pmatrix}1 & 1\\ -1 & -1\end{pmatrix}$. The scattering matrix product has now been fully specified and can be simplified using (23) for the leg chain prior to $j$ and for that after $j+1$. There remains the modified Gaussian integral $\int s_j s_{j+1}\exp[ik(R-L)]ds_1\ldots ds_n = \sqrt{}/ikr_j$ where $\sqrt{}$ stands for (25) (evaluated using the fact that the $j,j+1$ element of the inverse of the matrix associated with the quadratic form in (22) is $-1/r_j$). In summary for a diffracted ray with a surface leg the Green function is the zero $s$ value of the integrand of (21) with $F_j$ and $F_{j+1}$ replaced by $\dot{F}_j$ and $\dot{F}_{j+1}$, multiplied by $\sqrt{}/ikr_j$.

**Appendix C. Slit diffraction**

Slit diffraction in optics was first solved exactly by a scattering series due to Schwarzschild (of relativity fame) in 1902 [9],[11],[13]. Furthermore Schwarzschild proved that his scattering series was convergent (albeit slowly for long wavelength). (Complementing this in 1908, slit diffraction was solved exactly in elliptical coordinates by Sieger [14] as an infinite series of Mathieu functions which converges best for long wavelengths). It is of interest, then, to apply the present scattering series to the example of a single infinite straight line with a gap of width $a$ (the slit) in it. Each end of the gap is a $2\pi$ diffracting corner. The result will be that present series and the Schwarzschild series are identical term by term. The proven convergence in this special case gives some encouragement that the present series is convergent in general.

The geometry for the Schwarzschild solution is a plane wave (corresponding to a point source at infinite distance $r_\infty$) incident perpendicularly on the slit The boundary conditions on the screen are Dirichlet or Neumann ones but we restrict attention to Dirichlet. It suffices to verify that the wave field from corresponding scattering terms in the two series are identical on the line of the slit itself (at position *r*) since the field elsewhere can be deduced from this. Actually in the simple geometry of this example the wave field in the slit is the same as that produced by scattering in free space from two half unit magnetic flux lies (points) located at the gap ends so that no adaptation of the Stovicek scattering series is needed.

For the single scatter term only one end of the gap is relevant, the other half line can be removed, reducing to the Sommerfeld half line problem. The line end is taken as the origin. The two versions of the exact Sommerfeld solution are, then, on the left the present (Stovicek) expression with the Hankel function written as its asymptotic form for large argument (since $r_\infty$ is infinite), and on the right the Schwarzschild expression. The angles which have been substituted in (9) are $\phi=\pi/2$ and $\phi'=\pi$.

$$\int_{-\infty}^{\infty} ds_1 \frac{\exp\left[ik\sqrt{r_\infty^2 + r^2 + 2rr_\infty \cosh(s_1)}\right]}{\left(r_\infty^2 + r^2 + 2rr_\infty \cosh(s_1)\right)^{1/4}} \frac{1}{2\cosh(\tfrac{1}{2}(s_1 - i\tfrac{\pi}{2}))} = \frac{\exp(ikr_\infty)}{\sqrt{r_\infty}} \int_0^\infty dl_1 \sqrt{\frac{r}{l_1}} \frac{\exp[ik(l_1 + r)]}{l_1 + r} \quad (26)$$

The equality is verified by using $r_\infty=\infty$ to simplify the first term to $\exp(ikr_\infty)\exp[ikr\cosh(s_1)]/\sqrt{r_\infty}$, splitting the $(-\infty,\infty)$ domain of integration on the left into halves, and applying the change of coordinates $\cosh(s_1)=1+(l_1/r)$,. In the Schwarzschild expression on the right the scatter is pictured as happening not at the end of the half line but at distance $l_1$ along it, all such positions being integrated over (fig 13).

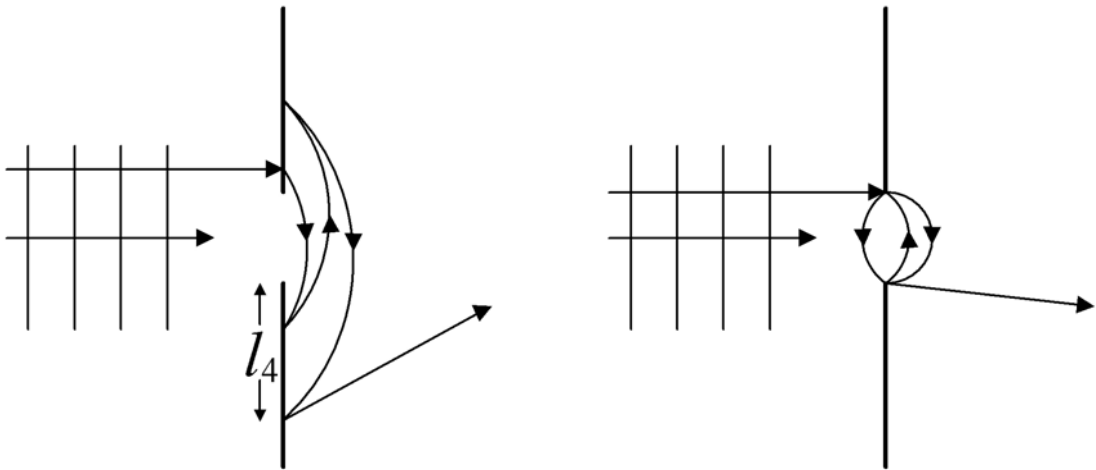

*Fig 13. Schwarzschild (left) versus the present (right) scattering solutions for slit diffraction. For the calculation in the text the observation position is taken as lying in the slit aperture (without loss of generality).*

For more than one scatter (the ones after the first having $\phi=\pi$, $\phi'=\pi$) the corresponding equality to be verified is

$$\int_{-\infty}^{\infty} ds_1...ds_n \ \exp[ik\{a\cosh(s_1) + ...a\cosh(s_1+...s_{n-1}) + r\cosh(s_1+..s_n)\}] \frac{1}{2\cosh(\frac{1}{2}(s_1 - i\frac{\pi}{2}))} \times$$

$$\left[\frac{1}{2\cosh(\frac{1}{2}s_2)}\right]\left[\frac{1}{2\cosh(\frac{1}{2}s_3)}\right]...\left[\frac{1}{2\cosh(\frac{1}{2}s_n)}\right]$$

$$= \int_0^{\infty} dl_1 dl_2...dl_n \ \sqrt{\frac{r(l_2+a)(l_3+a)...(l_n+a)}{l_1 l_2...l_n}} \ \frac{\exp[ik\{l_1 + (n-1)a + 2(l_2+...+l_n) + r\}]}{(l_1+a+l_2)(l_2+a+l_3)...(l_{n-1}+a+l_n)(l_n+r)}$$

(27)

The required change of coordinates to verify this is rather involved. First the integration variables on the left can be taken as $s_1$, $s_1+s_2$, ..., $s_1+s_2+...+s_n$. Then the (-

∞,∞) domain of integration on the left can be split into halves and converted to one with $s_1>0$, $s_1+s_2>0$, …, $s_1+s_2+…+s_n>0$. Define $l'_1 \equiv r[\cosh(s_1+s_2+..s_n)-1]$ and then $l'_2 \equiv (a/2)[\cosh(s_1+s_2+..s_{n-1})-1]$, $l'_3 \equiv (a/2)[\cosh(s_1+s_2+..s_{n-2})-1]$, …. $l'_n \equiv (a/2)[\cosh(s_1)-1]$. Then for the $n$ scatter term one has a form identical to the right side above except for the first and last factors in the last denominator

$$\int_0^\infty dl'_1...dl'_n \sqrt{\frac{r(l'_2+a)(l'_3+a)...(l'_n+a)}{l'_1 l'_2 ... l'_n}} \frac{2a \exp[ik\{l'_1+(n-1)a+2(l'_2+...+l'_n)+r\}]}{(2ar+2rl'_2+al'_1)(l'_2+a+l'_3)...(l'_{n-1}+a+l'_n)(a+2l'_n)}$$

(28)

Finally the coordinate change from $l'$ coordinates to $l$ coordinates is as follows. For the $n$ scatter term the first coordinate $l'_1$ is given by

$$l'_1 = l_n(2 + \frac{l_{n-1}}{l_n+a}(2 + \frac{l_{n-2}}{l_{n-1}+a}(2 + ...... \frac{l_2}{l_3+a}(2 + \frac{l_1}{l_2+a})))..) \qquad (29)$$

and the others, $2 \leq m \leq n$, by reduction to the first term for a lower number of scatters (in the square bracket) are given by

$$l'_m = \frac{a}{2(l_{n+2-m}+a)}\left[l_{n+1-m}(2 + \frac{l_{n-m}}{l_{n+1-m}+a}(2 + ...... \frac{l_2}{l_3+a}(2 + \frac{l_1}{l_2+a})))..)\right] \qquad (30)$$

This coordinate change converts the square root in the integral on the left directly to the square root in the integral on the right. The Jacobian of the transformation then equals the ratio of the remaining sections of integrand on the two sides of the equation. Both these assertions can be proved with some effort by induction – if they are true for the $n$-1 scatter then they are also true for the $n$ scatter. For example the Jacobian for the former is that for the latter times $(-1)^n (l_n+a)/(l_n \partial l_2'/\partial l_n - \frac{1}{2} a \partial l_1'/\partial l_n)$, which is easy to evaluate.

**Appendix D. Stovicek theory**

The justification for the concatenation formula (21) (or it predecessors) is most easily expressed in the multi-helicoid surface (officially the 'covering space') in which there is a Riemann helicoid at each scattering point. The helicoids do not have any unexpected connections - there is only one topology of path between any given source point and observation point – no alternative ones. All the different paths between them are deformable into each other within the surface. Two truths need to be verified for the justification of (21); that the formula obeys the wave equation as a function of observation position (except on shadow boundary lines), and that the discontinuities produced along shadow boundary lines are cancelled by others from separate terms.

That the concatenated formula obeys the wave equation follows by recognising that the quantities relating to the final leg to the observation point enter the formula just as they did for single scatter – the Sommerfeld solution. Therefore the wave equation is automatically obeyed.

As for the discontinuities, the cancellation comes about because for every line of discontinuity associated with *n* scatters there is a coincident line of discontinuity associated with *n*+1 scatters. An example is shown in fig 14. The cancellation is guaranteed by the same pole crossing argument as for the single scatter Sommerfeld solution.

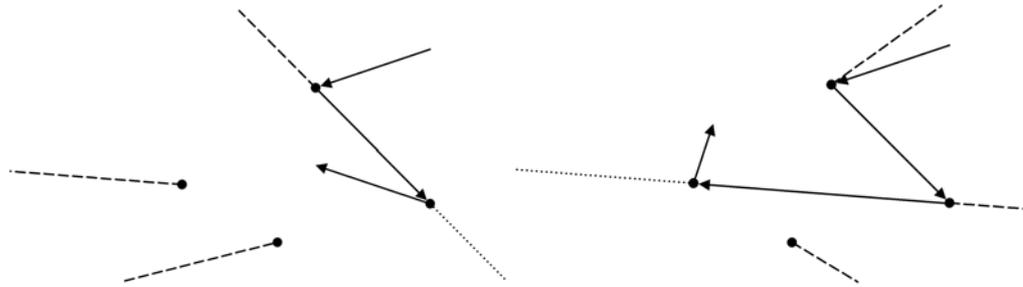

*Fig 14. Left: The four discontinuity lines associated with a particular double scatter among four points. Three go radially away from the last scatter, and one, based at the last scatter goes radially away from the penultimate scatter. Right: The four discontinuity lines associated with a particular triple scatter among the same points. They are generated as before, and one of them, the leftmost coincides with (and cancels) the double scatter discontinuity.*

Acknowledgements
We thank O. Giraud for his detailed reading and corrections. The funding of the Leverhulme Trust for A. Thain is gratefully acknowledged. JHH is grateful to KG Budden for extolling to his class thirty years ago the virtues of complex angle in wave propagation.